# Distortion-Corrected Image Reconstruction with Deep Learning on an MRI-Linac

Shanshan Shan[1,2,3*], Yang Gao[3*], Paul Z. Y. Liu[1,2], Brendan Whelan[1,2], Hongfu Sun[3], Bin Dong[2], Feng Liu[3], David E. J. Waddington[1,2]

[1]ACRF Image X Institute, Sydney School of Health Sciences, Faculty of Medicine and Health, The University of Sydney, Sydney, New South Wales, Australia.

[2]Department of Medical Physics, Ingham Institute of Applied Medical Research, Liverpool, New South Wales, Australia.

[3]School of Information Technology and Electrical Engineering, The University of Queensland, Brisbane, Queensland, Australia.

*Shanshan Shan and Yang Gao contribute equally to this work. Please direct correspondence to shanshan.shan@sydney.edu.au.

Total word count: 5633

**Abstract**

**Purpose:** Magnetic resonance imaging (MRI) is increasingly utilized for image-guided radiotherapy due to its outstanding soft-tissue contrast and lack of ionizing radiation. However, geometric distortions caused by gradient nonlinearities (GNLs) limit anatomical accuracy, potentially compromising the quality of tumour treatments. In addition, slow MR acquisition and reconstruction limit the potential for effective image guidance. Here, we demonstrate a deep learning-based method that rapidly reconstructs distortion-corrected images from raw k-space data for MR-guided radiotherapy applications.

**Methods:** We leverage recent advances in interpretable unrolling networks to develop a Distortion-Corrected Reconstruction Network (DCReconNet) that applies convolutional neural networks (CNNs) to learn effective regularizations and nonuniform fast Fourier transforms for GNL-encoding. DCReconNet was trained on a public MR brain dataset from eleven healthy volunteers for fully sampled and accelerated techniques, including parallel imaging (PI) and compressed sensing (CS). The performance of DCReconNet was tested on phantom, brain, pelvis, and lung images acquired on a 1.0T MRI-Linac. The DCReconNet, CS-, PI-and UNet-based reconstructed image quality was measured by structural similarity (SSIM) and root-mean-squared error (RMSE) for numerical comparisons. The computation time and residual distortion for each method were also reported.

**Results:** Imaging results demonstrated that DCReconNet better preserves image structures compared to CS- and PI-based reconstruction methods. DCReconNet resulted in the highest SSIM (0.95 median value) and lowest RMSE (<0.04) on simulated brain images with four times acceleration. DCReconNet is over ten-times faster than iterative, regularized reconstruction methods.

**Conclusions:** DCReconNet provides fast and geometrically accurate image reconstruction and has the potential for MRI-guided radiotherapy applications.

**Keywords:** MRI-guided radiotherapy, Geometric distortion, Compressed sensing, Parallel imaging, Unrolling network

# 1. Introduction

Hybrid systems integrating an MRI scanner and a linear accelerator (Linac) have recently been developed to perform image-guided radiotherapy. Such MRI-Linacs have superior soft-tissue contrast compared to other imaging technologies, such as computed tomography (CT) and do not require additional imaging radiation exposure [1, 2]. These new treatment systems enable fast tumor imaging that facilitates adaptive dose delivery and enhanced conformal treatments [3-5]. An ongoing challenge in the development of MRI-Linac systems is the rapid reconstruction of images with high geometric accuracy. MR image distortion limits geometric accuracy and can cause target localization errors that risk degrading the quality of MRI-Linac tumor treatments [6, 7]. A leading source of image distortion on MRI-Linac systems is gradient nonlinearities (GNLs) that result from design compromises made to the MRI scanner to enable the integration of the X-ray subsystem [8].

In conventional MRI systems, gradient fields are assumed to vary linearly across the field of view (FOV), thus enabling the spatial encoding of MR signals [9]. However, generating spatially linear gradient fields is not as easily achievable on MRI-Linac systems due to engineering constraints on gradient coil design and manufacture [10]. Further, gradient linearity may be intentionally compromised for increased slew rate, higher gradient amplitudes and reduced peripheral nerve stimulation [11, 12]. In addition, eddy currents caused by fast-switching gradients during the image encoding process can also deviate actual gradient fields [13]. MR reconstruction methods (e.g., inverse Fourier transform) operate under the assumption that the acquired dataset has been encoded using linear gradients [14]. The presence of GNL undermines the gradient encoding and causes geometric distortions in reconstructed MR images, which is problematic for applications requiring high geometric fidelity, such as MR-guided radiotherapy (MRIgRT) [15-17].

Standard approaches to GNL correction utilize prior knowledge of the GNL field to correct geometric deformation after MR image reconstruction using coordinate mapping and intensity scaling [18]. Image-based interpolation is usually applied to approximate the coordinate mapping operation because of pixel shrinkage and/or dilation [19, 20]. These techniques are widely adopted on most commercial MR scanners; however, studies have shown that such image-based distortion-correction methods have an intrinsic smoothing effect that results in signal blurring and resolution loss, especially at the edges of large FOVs [21, 22]. While interpolation-based distortion correction is sufficient to delineate structures of interest near the center of the FOV, MRI-Linac treatments can have targets positioned off-axis due to the limited degrees of freedom and limited range of motion of the MRI-Linac patient couch [23]. Besides, studies of distortion on low-field MRI-Linacs have found that further GNL correction should be considered for targets greater than 50 mm from the isocenter [24] and is explicitly recommended beyond a radius of 100 mm [25]. It is also noted that precise distortion correction across the entire FOV enables the implementation of MR-only treatment planning on the MRI-Linac. Such MR-only treatment planning allows treatments to be simulated and delivered on the same system. Still, this workflow relies on the generation of geometrically and dosimetrically accurate synthetic CTs that encompass the whole body [26], and peripheral distortion is a limiting factor in the generation of synthetic CT scans from MRI data [27].

To address the limitations of image interpolation-based distortion correction, Tao *et al.* proposed a k-space domain-based method to prospectively correct GNL distortions during the image reconstruction process [28, 29]. This method incorporates GNL with gradient encoding to form an ill-posed problem that, when solved, reconstructs distortion-corrected (DC) images directly from the k-space domain. Typically, optimization algorithms with regularizations (e.g., wavelet, total variation, and low rank) are used to solve this ill-posed problem. However, it is challenging to determine the optimal regularization parameters, and the algorithms are computationally expensive [30], making them impractical for image guidance in clinical practice.

Deep learning has shown great potential to solve optimization-based MR reconstruction problems, avoiding cumbersome parameter-tuning processes and performing fast online reconstruction [31, 32]. Recently, we developed a ResUNet-based network [33] to learn the relationship between undistorted and distorted brain images, and imaging results showed that it could correct geometric distortions successfully. However, this method is limited to fully sampled images. Transfer learning is required for other imaging targets that have never been seen during the training process (e.g., pelvis and phantom data), making this method less generalizable. Here, we develop and investigate a general distortion-corrected reconstruction neural network (DCReconNet) for fast imaging on an MRI-Linac with fully sampled and undersampled acquisitions. Based on an interpretable unrolling network architecture [34, 35], DCReconNet uses convolutional neural networks (CNN) to learn effective regularizations and nonuniform fast Fourier transforms (NUFFT) for GNL-encoding operation. MR acceleration techniques, including compressed sensing (CS) [36] and parallel imaging (PI) [37] were incorporated into DCReconNet to further reduce MR acquisition time. The proposed network was trained on a public MR brain dataset from eleven healthy volunteers, and the performance was compared with our previously developed ResUNet-based method and conventional regularization-based methods on phantom, human brain images, pelvis images and patient lung images acquired from an MRI-Linac for fully sampled and retrospectively subsampled acquisitions.

## 3. Methods

### 3.1 GNL-encoding model and inverse reconstruction

The forward gradient encoding with GNL can be formulated as [28, 29]:

$$mE_{GNL}°Ax = b \qquad (1)$$

Where $b$ denotes the measured GNL-corrupted k-space data, and $x$ is the distortion-corrected image to be reconstructed. $m$ represents the undersampling matrix calculated from a specific sub-sampling mask. For single channel acquisitions, $A = F$ and $F$ denotes the theoretical Fourier transform matrix with the kernel of $e_{k,L} = e^{-2\pi jkL}$, where $L$ is the theoretical encoding position. For multi-channel acquisitions, $A = FS$ and $S$ is the coil sensitivity [28]. $E_{GNL}$ is the

GNL-encoding operator with the form of $e_{k,\Delta(L)} = e^{-2\pi jk\Delta(L)}$ and $\Delta(L)$ represents the GNL-induced spatial deviation at theoretical location $L$. $E_{GNL} \circ A$ is the element-wise multiplication of matrix $E_{GNL}$ and $A$. Defining the forward matrix operation $\Phi = mE_{GNL} \circ A$; it represents a nonuniform to uniform spatial mapping for Cartesian sampling, which can be implemented by Type-I nonuniform fast Fourier transform (NUFFT) [38]. Eq. (1) describes an ill-posed problem that can be solved by the following equation:

$$x = \underset{x}{\mathrm{argmin}}\{\|\Phi x - b\|_2^2 + \lambda R(x)\} \tag{2}$$

where $\|\Phi x - b\|_2^2$ represents the L2 norm vector that promotes the data fidelity between estimated and measured data. $R(x)$ is the sparsifying regularization function (e.g., wavelet, total variation, and low rank) with weighting parameter $\lambda$ [39, 40]. Conventional regularization methods have typically aimed to maximize the accuracy of the magnitude of the reconstructed image. Recent studies have shown that phase regularizations, such as CS with phase cycling, enable better image reconstruction than conventional methods [41, 42]. However, the performance of regularization methods is determined by the weighting $\lambda$, which requires careful tuning to find the optimal one. In addition, iterative algorithms (e.g., non-linear conjugate gradient) [36] are typically used to solve regularization problems, which are computationally expensive and unsuitable for clinical translation.

**3.2 Network architecture**

In this work, we developed a deep learning-based method to solve the ill-posed problem in Eq. (1) and to reconstruct distortion-corrected images directly from the k-space domain. The developed network takes two inputs: 1) the acquired GNL-corrupted k-space with a size of 256 × 256, and 2) the corresponding GNL field along in-plane directions (i.e., x- and y-directions for axial planes) with a size of 256 × 256 × 2. The GNL field information was provided by our previously developed GNLNet [43]. The output of DCReconNet is the distortion-corrected image.

As an alternative to slow iterative algorithms, model-driven unrolling networks have been widely used to provide a robust and rapid solution to MR reconstruction, incorporating known MR physics and having well-defined interpretability [34]. Based on an unrolling network architecture [34], Eq. (1) can be solved by the equation below:

$$x = \underset{x}{\mathrm{argmin}}\{\|\Phi^T \Phi x - \Phi^T b\|_2^2 + CNN(x)\} \tag{3}$$

DCReconNet was composed of *N=7* iterative soft shrinkage-thresholding blocks, as shown in Figure 1. Each block starts with a data fidelity term $\Phi^T \Phi x - \Phi^T b$ with GNL-encoding operation, which computes a residual in k-space and then projects it back to the image domain. Then forward and backward nonlinear transforms with a soft-thresholding operation are used to reduce image artefacts. Each nonlinear transform includes two linear convolutional operators split by a rectified linear unit (ReLu). A skip connection is added to form a residual block and to further facilitate network training. A data consistency layer is added at the end of each block (except for the last block) to enforce consistency in k-space, which means the reconstructed k-space data at the sampled locations is replaced by the original acquired data. The data consistency step was removed for multi-channel reconstructions to reduce the computational cost, and only the magnitude images were involved in these calculations.

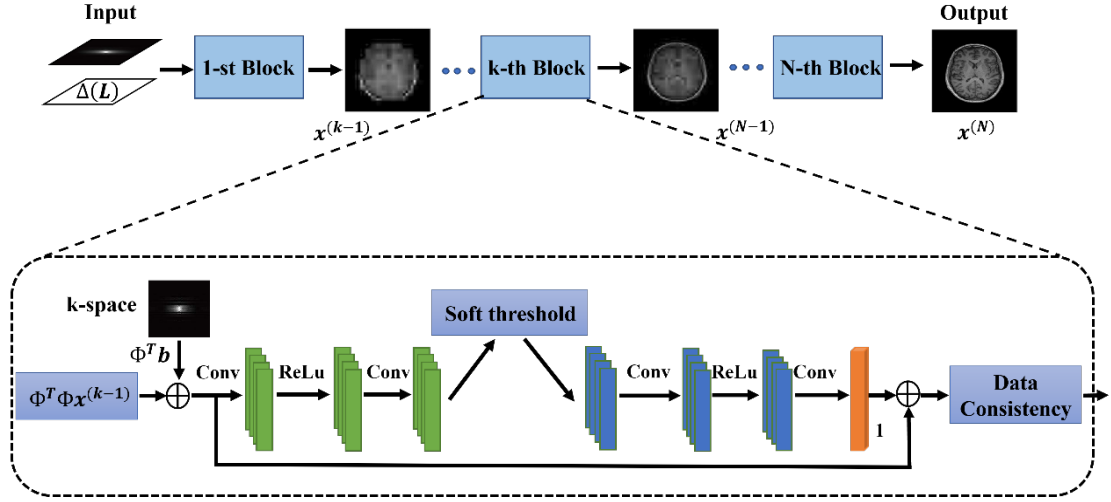

Figure 1. The DCReconNet architecture is composed of $N$ iterative soft shrinkage-thresholding blocks. Each block starts with a data fidelity calculation, followed by nonlinear transforms and a soft thresholding operation. The nonlinear transforms include forward and backward transforms, and each is designed as the combination of a rectified linear unit (ReLU) and two linear convolutional operators. A skip connection is used to form a residual block, and a data consistency operation is added to enforce k-space consistency with measured k-space samples at the end of each block.

**3.3 Data preparation for network training**

A public T1-weighted brain dataset [44] was used to generate training data for the proposed network. Brain images from eleven healthy volunteers were acquired with a whole-body MRI scanner (Magnetom Tim Trio; Siemens Healthcare, Germany). 300 slices covering the entire brain volume were selected from each volunteer, and 300 × 11 = 3300 brain images were used as labels (ground truth). The imaging parameters are as follows: image size= 320 × 320 × 256, resolution=0.7 mm × 0.7 mm × 0.7 mm, and TE/TR = 2.13 ms/2.4 s. All brain images were cropped into the same matrix size of 256 × 256. To cover the entire region of interest (ROI) of 30 cm × 30 cm × 30 cm (shown in Figure 2(a)), twenty axial planes ranging from -150 mm to 150 mm were allocated with equal intervals (15mm) along the z direction. Repeated operations were performed on coronal (along y direction) and sagittal planes (along x direction), respectively. In total, sixty orthogonal planes were determined, and fifty brain images randomly selected from the label dataset were positioned at each plane. Thus, 60 × 50 = 3000 brain images from ten volunteers served as training data, and the 300 remaining brain images from the eleventh volunteer were used to prepare the simulated testing data.

To simulate GNL-corrupted k-space data, our previously developed GNLNet [43] was used to provide the GNL field information for the forward GNL-encoding operation in Eq. (1). These simulated GNL-corrupted k-space datasets were retrospectively subsampled using 1D random sampling mask along the phase encoding direction with acceleration factors (AF) of 2, 4 and 6. The fully sampled and subsampled k-space data with label images were then fed into DCReconNet as training datasets (Figure 2(b)).

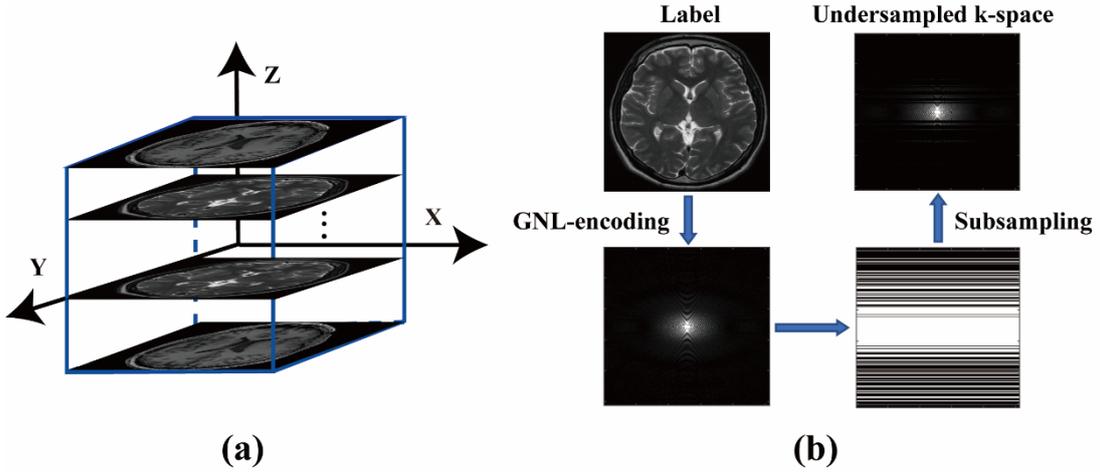

Figure 2. (a) The ROI is 30 cm × 30 cm × 30 cm. Twenty uniformly-spaced axial planes were allocated, ranging from -150mm to 150mm along the z direction in the ROI. Repeated operations were performed on coronal (along y direction) and sagittal planes (along x direction), respectively. (b) The GNL-encoding operation was used on the label images to simulate GNL-corrupted k-space data. 1D random sampling mask with AF of 2 was retrospectively applied to subsample the simulated k-space data. Training data preparation was repeated for AFs of 4 and 6.

**3.4 Data acquisition for model testing**

To test our network, a 3D distortion phantom with precisely known geometric properties [37] was scanned with a body coil on the Australian MRI-Linac system. The imaging parameters were turbo spin echo sequence (TSE), image size = 130 × 110 × 192, resolution = 1.8 mm × 2 mm × 1.8 mm, image bandwidth = 202 Hz, TE/TR = 15 ms/5.1 s, and phase encoding direction: R/L. The acquired phantom data was fully sampled on the scanner and then retrospectively undersampled at AF=4 using a 1D random subsampling mask.

To evaluate the performance of the proposed DCReconNet on multi-channel acquisitions, a fully sampled head phantom scan was conducted with a 6-channel head coil and the following imaging parameters: TSE sequence, image size = 256 × 256 × 40, resolution = 0.98 mm × 0.98 mm × 5 mm, image bandwidth = 203 Hz, TE/TR = 101 ms/12 s, and phase encoding direction: R/L. The fully sampled brain phantom data were retrospectively subsampled with AFs of 2 and

4. The head phantom was also scanned with a CT system (Philips, Brilliance Big Bore) and the imaging parameters: image size= 512 × 512 × 325 and resolution = 0.78 mm × 0.78 mm × 1 mm. These CT images were used as (undistorted) reference images.

A healthy volunteer was scanned with a 6-channel head coil, and the acquisitions were retrospectively undersampled by AFs of 2 and 4 using the previously described masks. Two scans were conducted to cover the whole ROI; the brain center was consistent with the scanner isocenter for the first scan, and the brain center was shifted by 5 cm along x direction for the second scan. The scan parameters were as follows: image size = 256 × 256 × 12, resolution = 0.98 mm × 0.98 mm × 5 mm, image bandwidth = 203 Hz, TE/TR = 77 ms/8 s, and phase encoding direction: R/L. A patient with a lung tumor was scanned with an 8-channel torso coil, and the acquisitions were retrospectively undersampled by AFs of 2 and 4. The scan parameters were as follows: image size = 256 × 256 × 10, resolution = 1.56 mm × 1.56 mm × 4 mm, image bandwidth = 201 Hz, TE/TR = 12 ms/0.8 s, and phase encoding direction: R/L. The Sum-of-Squares (SoS) of the zero-filled reconstructed images from all channel k-space data were used as inputs to the network so that multi-channel data could be reconstructed without needing coil sensitivity maps [45]. Pelvis images from a healthy male volunteer were acquired with a body coil and were retrospectively undersampled by AFs of 2 and 4. The imaging parameters were: image size = 154 × 192 × 44, resolution = 2.1 mm × 2.1 mm × 3.5 mm, image bandwidth = 202 Hz, TE/TR = 86 ms/10.2 s, and phase encoding direction: R/L. A clinical 3T MRI scanner (Siemens Skyra with 50cm DSV) with vendor's correction was also used to scan the pelvis with the imaging parameters: image size = 256 × 208 × 51, resolution = 1.8 mm × 1.8 mm × 3.5 mm, image bandwidth = 400 Hz, TE/TR = 86 ms/10.2 s, and phase encoding direction: A/P. The vendor's correction on the 3T clinical scanner has been verified as geometrically accurate within the FOV [46] and therefore 3T corrected images were used as references.

### 3.5 Network training

Mean square error (MSE) was used to calculate the loss function during network training. The training process utilized the Adam optimizer [47] and a batch size of 32. Networks were trained for 100 epochs with learning rates of 0.001 and 0.0001 for the first 50 and the remaining training epochs, respectively. The DCReconNet was trained on a high-performance computer equipped with an Nvidia Tesla V100 P32 GPU, and the training took ~20 hours for 100 epochs. Human studies were conducted with the approval of the Institutional Review Board (IRB). The source code for our neural networks is available at: https://github.com/shanshanshan3/DCReconNet.

### 3.6 Evaluation

In this study, the CS phase cycling method [42] was implemented to solve distortion-corrected image reconstruction problems in Eq. (3) for single-channel and multi-channel PI acquisitions, referred to as DCCS and DCCS-PI, respectively. NUFFT was used for distortion-corrected zero-filling reconstruction, referred to as DCZF. In addition to conventional regularization methods, a sequence of two networks (unrolling and ResUNet) was also implemented, referred to as UnUNet. The UnUNet includes a standard unrolling network to reconstruct distorted images from undersampled k-space data and a ResUNet to correct image distortion. The UnUNet was trained on the same brain dataset described in section 3.3. The proposed DCReconNet was compared with DCZF, DCCS, DCCS-PI and UnUNet for undersampled acquisitions with AFs=2, 4 and 6. For the DCCS-PI method, coil sensitivity maps were calculated from the k-space center (matrix size 24 × 24) using ESPIRiT [48]. Regularization parameters were selected for each sampling mask via a grid search that minimized MSE between reconstructed and reference images. The conventional Fourier transform method and the proposed DCReconNet were also used for image reconstruction with fully sampled acquisitions, which are referred to as FT and DCReconNet-FS. To quantitatively evaluate the quality of reconstructed images, the root mean square error (RMSE) and structural similarity index (SSIM) were calculated between reconstructed and reference images. A total of 3718 marker positions were extracted from the 3D distortion phantom to quantitatively measure the geometric distortion before and after correction.

## 4. Results

4.1 Simulation results

Brain image reconstruction results from the testing dataset (300 brain slices described in section 3.3) are shown in Figure 3. The GNL-corrupted k-space data were simulated at an axial plane (z=120 mm), a coronal plane (y=33 mm) and a sagittal plane (x=130 mm), respectively. Yellow lines represent contours of ground truth brain images. The contours of FT-reconstructed images do not match with yellow lines, indicating geometric deformation, including image shrinkage and dilation. A subsampling mask with AF=4 was imposed on the GNL-corrupted k-space data and then reconstructed by DCZF, DCCS, UnUNet and DCReconNet methods. As indicated by the yellow lines, geometric distortions were successfully corrected in images reconstructed by these four methods. Severe artefacts and noise are presented on DCZF-reconstructed images, and these artefacts were substantially reduced when using the DCCS, UnUNet and DCReconNet algorithms. Compared with DCCS and UnUNet methods, DCReconNet resulted in lower-level error maps with smaller RMSE and higher SSIM values, suggesting better image quality.

RMSE and SSIM values were calculated across the 300 test images reconstructed by DCZF, DCCS, UnUNet and DCReconNet methods (see boxplot results in Figure 4). For the DCZF-reconstructed images, the median and maximum RMSE values are 0.05 and 0.09, respectively. The maximum RMSE values for the other three methods are lower than 0.04, demonstrating that more accurate reconstructions were achieved than the DCZF method. It is also noticeable that DCReconNet and UnUNet give lower RMSE values than the DCCS method. In terms of SSIM, images reconstructed by the proposed DCReconNet and UnUNet show the highest level (0.95 median value) than those reconstructed by DCZF (0.65 median value) and DCCS (0.86 median value) methods.

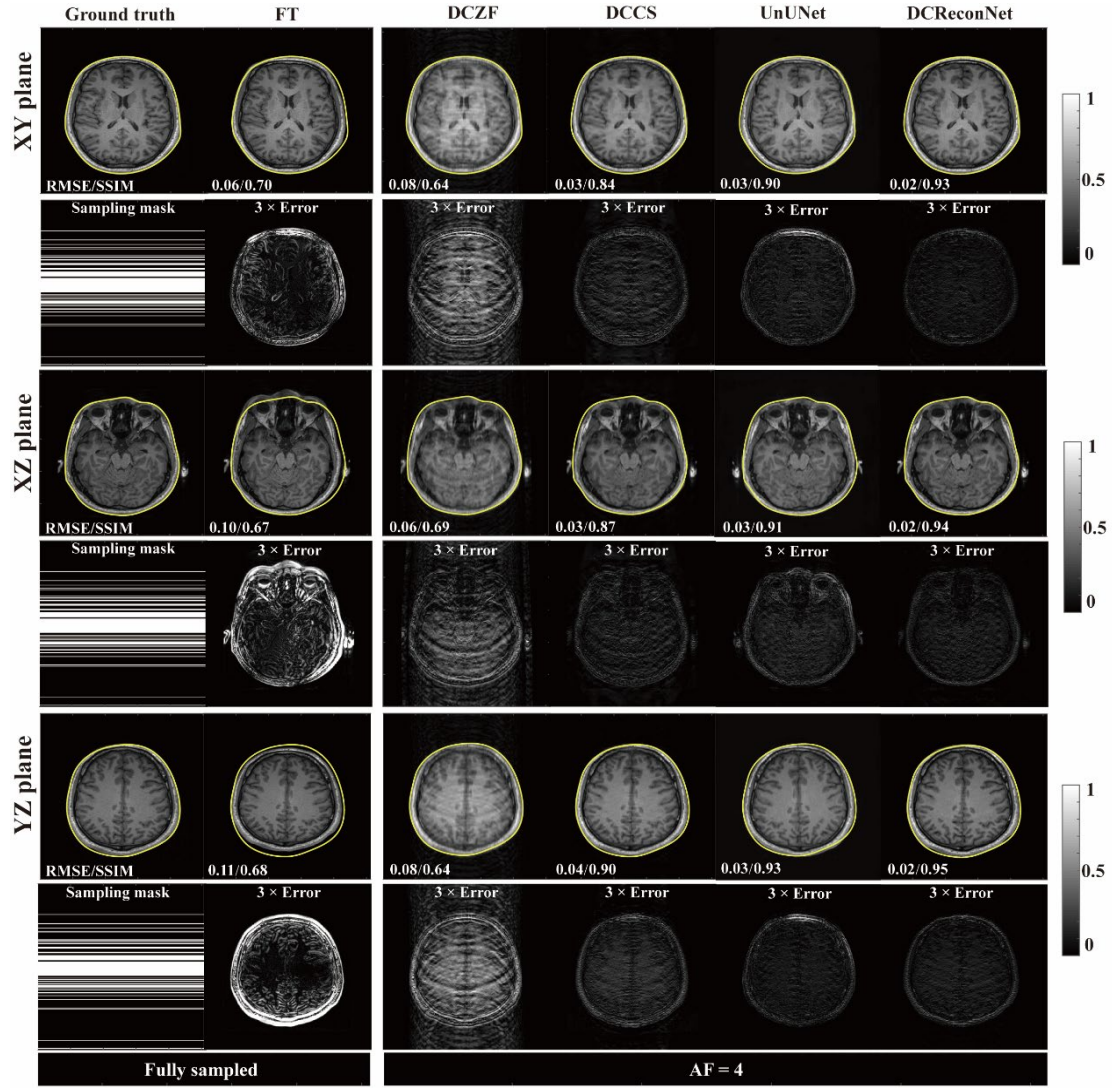

Figure 3. Comparison of different reconstruction methods on GNL-corrupted data simulated in three orthogonal planes for AF = 4. Images are reconstructed at the XY plane of z=120 mm, XZ plane of y=33 mm and YZ plane of x=130 mm, respectively. Ground truth images are shown at the first column, and the FT reconstructed images from fully sampled GNL-corrupted k-space data are shown inat the second column, followed by DCZF-, DCCS-, UnUNet- and DCReconNet-reconstructed images using retrospectively subsampled k-space. Three-folder error maps between FT, DCZF, DCCS, UnUNet and DCReconNet reconstructed images, and the ground truth images are shown at the bottom. RMSE and SSIM values were calculated and displayed at the bottom of figures. The yellow lines denote the contours of ground truth brain images.

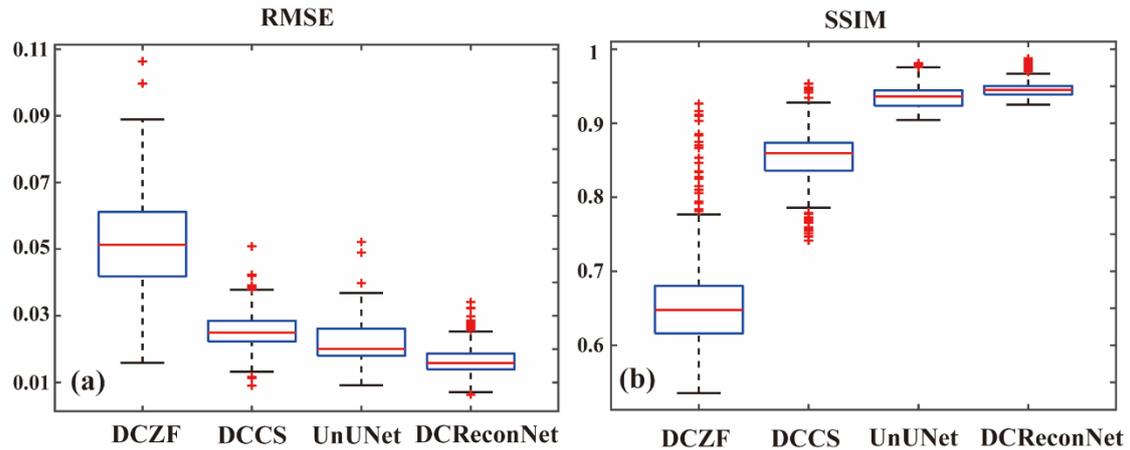

Figure 4. Boxplots of (a) RMSE and (b) SSIM values across 300 test images reconstructed by DCZF, DCCS, UnUNet and DCReconNet methods, respectively, with the AF = 4. Minimum, first quartile (25%), median (50%), third quartile (75%), and the maximum value were statistically evaluated. Red crosses denote outliers, accounting for 0.7% of total samples.

DCReconNet is compared to DCZF, DCCS and UnUNet methods for the reconstructions of simulated brain data with AFs=2, 4 and 6 in Figure 5. DCReconNet, UnUNet and DCCS reconstructed images with comparable quality at AF = 2. As pointed out by yellow arrows, image blurring and structural detail loss were observed on DCCS-reconstructed images at AF = 4 and AF = 6. These structural details were well preserved using DCReconNet and UnUNet, demonstrating the superior performance of the neural network approach to high AF subsampling cases. Similarly, the DCReconNet and UnUNet led to lower RMSE and higher SSIM than the DCCS method. Considerable geometric distortions are evident on fully sampled FT-reconstructed images; however, these distortions are almost entirely removed in the other images, as indicated by yellow contouring lines.

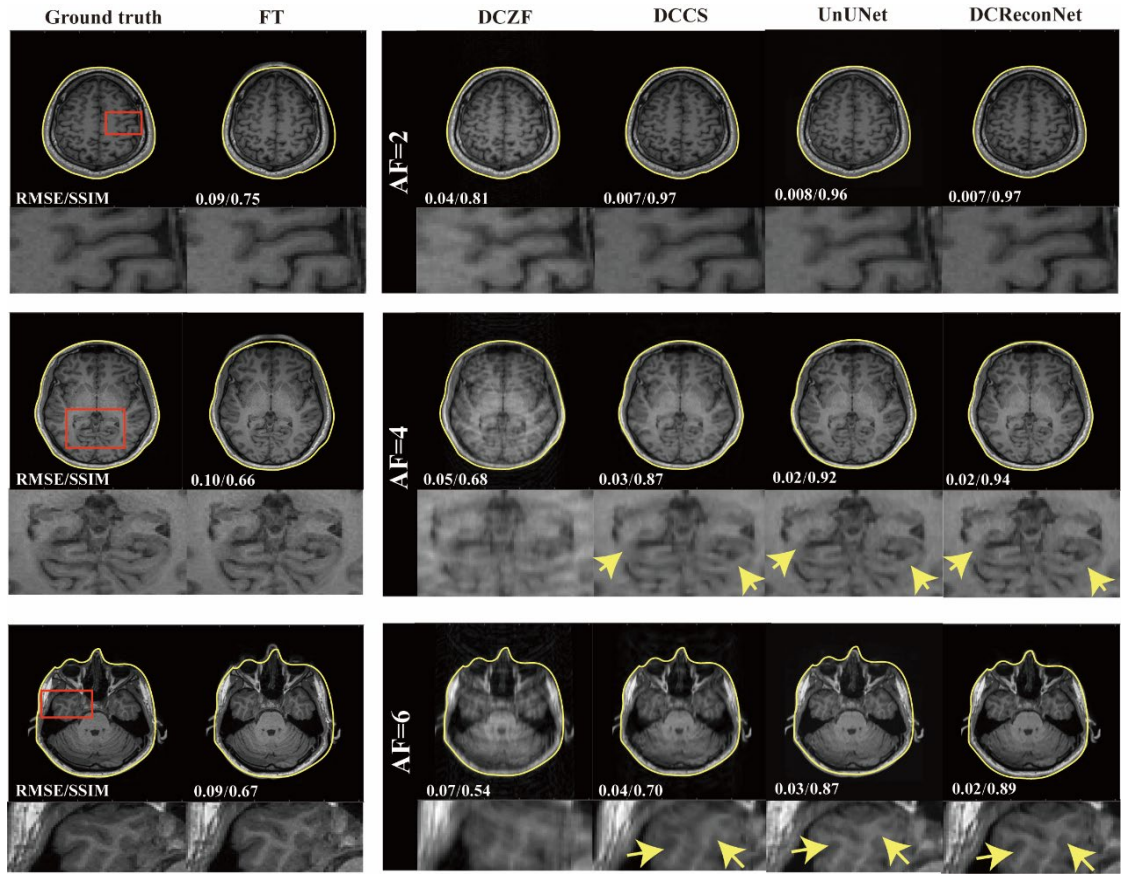

Figure 5. Reconstruction of brain images from simulated GNL-corrupted data at different acceleration factors. DCZF, DCCS, UnUNet and DCReconNet reconstructions are shown for an axial slice (z = 102 mm) with AF = 2, AF = 4 and AF = 6. Conventional FT was used to reconstruct fully sampled data. Zoomed regions (red rectangle) are shown for each AF. Yellow arrows indicate that structural details are better preserved by DCReconNet and UnUNet. Yellow lines are the contour boundary from the ground truth images and are included for comparison of image geometric distortion.

4.2 Experimental results

4.2.1 Phantom results

Images of a gridded distortion phantom were acquired with a body coil from the Australian MRI-Linac system to evaluate the performance of DCReconNet. Geometric distortions are present in fully sampled FT-reconstructed images (Figure 6 (a)), where straight grids are warped due to GNL. These distortions were significantly reduced on DCReconNet-FS images (Figure 6 (b)), which served as reference images for undersampling reconstructions. The reconstruction results of DCZF, DCCS, UnUNet and DCReconNet at AF=4 are shown in Figure 6 (c-f). Undesired artifacts were presented in UnUNet-reconstructed images (Figure 6 (e)) in comparison to DCReconNet results (Figure 6 (f)).

A total of 3718 marker positions extracted from the 3D distortion phantom were used to quantitatively measure the geometric deformation before and after correction using the

DCReconNet, as shown in Table 1. The DCReconNet reduced the maximal displacement within 2 mm, showing a dramatic decrease compared with uncorrected images (14.1 mm). Similarly, the RMSE of corrected marker positions (0.4 mm) is approximately one-tenth of uncorrected ones (3.4 mm). We also shifted the geometry phantom by 5cm along the axial dimension (x direction), and distortions on 3718 marker positions with and without correction were shown in Supporting Information Figure S4. Before correction, geometric distortions less than 1mm occur mostly in the central area. Markers with 1mm-3mm distortions can be partly seen in the central area and spread out towards the edges. After correction, image distortions on all marker positions are less than 2mm. Axial phantom images reconstructed by FT, DCZF, DCCS and DCReconNet methods are shown in Supporting Information Figure S5. As the images covered the edge of the DSV, larger geometric distortions are visible in FT-reconstructed images compared to Figure 6. The DCReconNet successfully reduced geometric distortions and better preserved structural details than DCCS and DCZF methods at AF = 4.

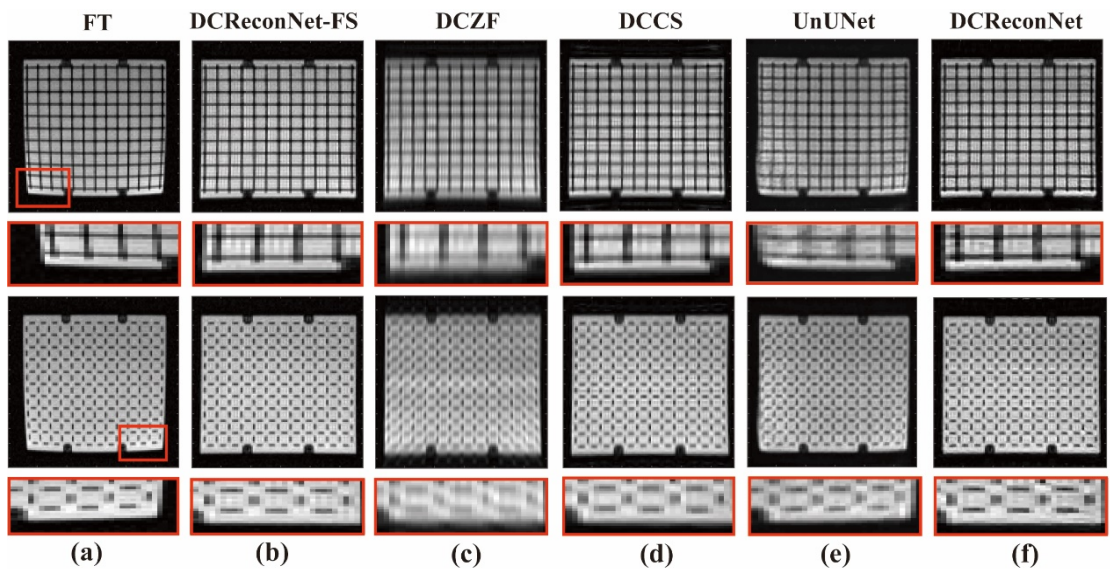

Figure 6. Grid phantom images acquired with a body coil from the MRI-Linac scanner at the location of z = 7.5 mm (the top row) and z = -26 mm (the bottom row), respectively. (a) fully sampled grid phantom images reconstructed by FT. (b) fully sampled images reconstructed by DCReconNet. (c)-(f) subsampled reconstructed images at AF = 4 using DCZF, DCCS, UnUNet and DCReconNet, respectively. Zoom-in areas show the phantom structure details.

Table 1 RMSE and maximal error of phantom markers before and after distortion correction

|  | Parameters | X (mm) | Y (mm) | Z (mm) | R (mm) |
|---|---|---|---|---|---|
| Uncorrected | Maximal error | 6.6 | 10.2 | 8.8 | 14.1 |
|  | RMSE | 1.6 | 2.2 | 2.1 | 3.4 |
| Corrected | Maximal error | 1.1 | 0.9 | 1.3 | 1.5 |
|  | RMSE | 0.2 | 0.2 | 0.3 | 0.4 |

Brain phantom images acquired on our MRI-Linac with a 6-channel head coil are shown in Figure 7. Yellow lines denote the contours of reference (undistorted) CT images. The FT-reconstructed images do not align with yellow lines because of geometric distortion. After applying the DCReconNet-FS, the images align better with CT contours, as indicated by red arrows. For the undersampled acquisition with AF = 4, image metrics (SSIM/RMSE) showed that DCReconNet (0.82/0.01) gave better reconstructions than the DCCS-PI (0.78/0.02) and the UnUNet (0.69/0.04) methods. Compared with the UnUNet-reconstructed image at AF=4, less artifacts were seen on the DCReconNet-reconstructed image, as pointed by the orange arrows.

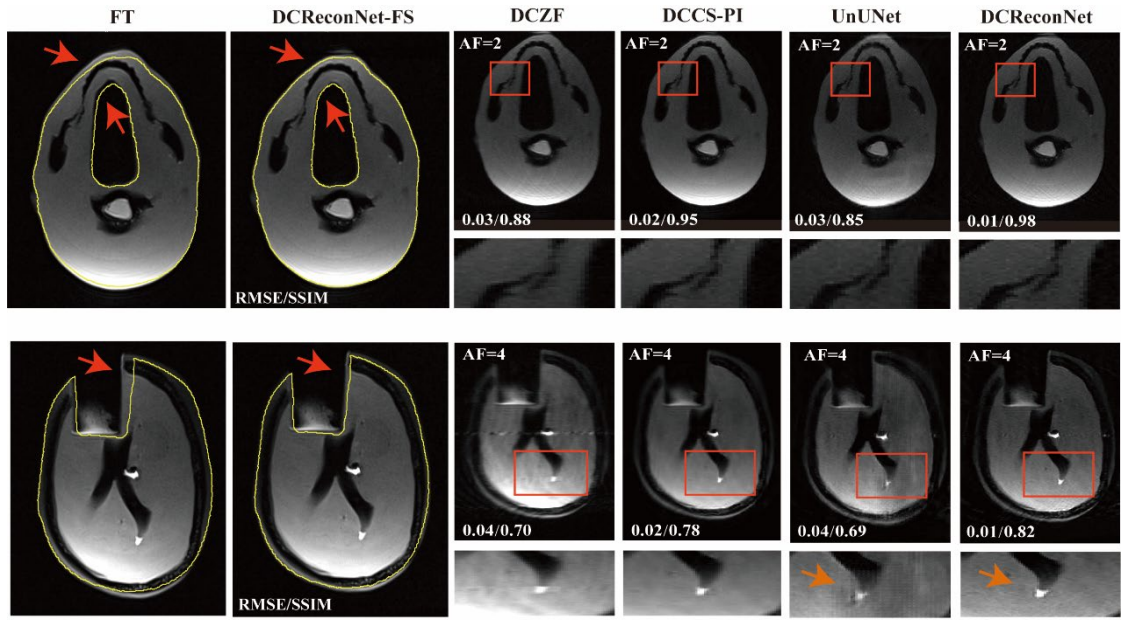

Figure 7. Brain phantom images reconstructed with AF = 2 and AF =4 at slice positions of y= 17.5 mm (the top row) and y=22.5 mm (the bottom row). Yellow lines represent the contours of reference CT brain phantom images. Zoom-in areas show image structural details.

4.2.2 Volunteer brain, pelvis and patient lung results

Volunteer brain images acquired with a 6-channel head coil are shown in Figure 8. For AF = 2, DCReconNet (Figure 8 (e)) and DCCS-PI (Figure 8 (d)) resulted comparable reconstructions with similar RMSE (0.01) and SSIM (0.92) values, which is consistent with results presented in Figure 5 and Figure 7. For AF = 4, image structural details were effectively preserved in Figure 8 (j) in comparison with Figure 8 (i), indicating that the DCReconNet led to better reconstruction performance than DCCS-PI method. Patient images with a lung tumor acquired using an 8-channel torso coil are shown in Figure 9 (a). Comparable results were achieved with DCReconNet and DCCS-PI methods for AF=2. Image structural details are shaper in DCReconNet-reconstructed images (0.02/0.88) with higher SSIM and lower RMSE than the DCCS-reconstructed images (0.03/0.84) for AF=4. Pelvis images of a healthy volunteer acquired from a clinical 3T MRI scanner with vendor's correction were shown in Figure 9 (b) and were used as reference images. Reconstruction methods of FT, DCCS and DCReconNet were compared on pelvis images acquired from our MRI-Linac scanner with a body coil. Significant distortions are shown in FT-reconstructed images, particularly at the edge of FOV, as indicated by the red arrows. The DCCS and DCReconNet methods reduced image distortions, and better structural details are seen in DCReconNet-reconstructed images.

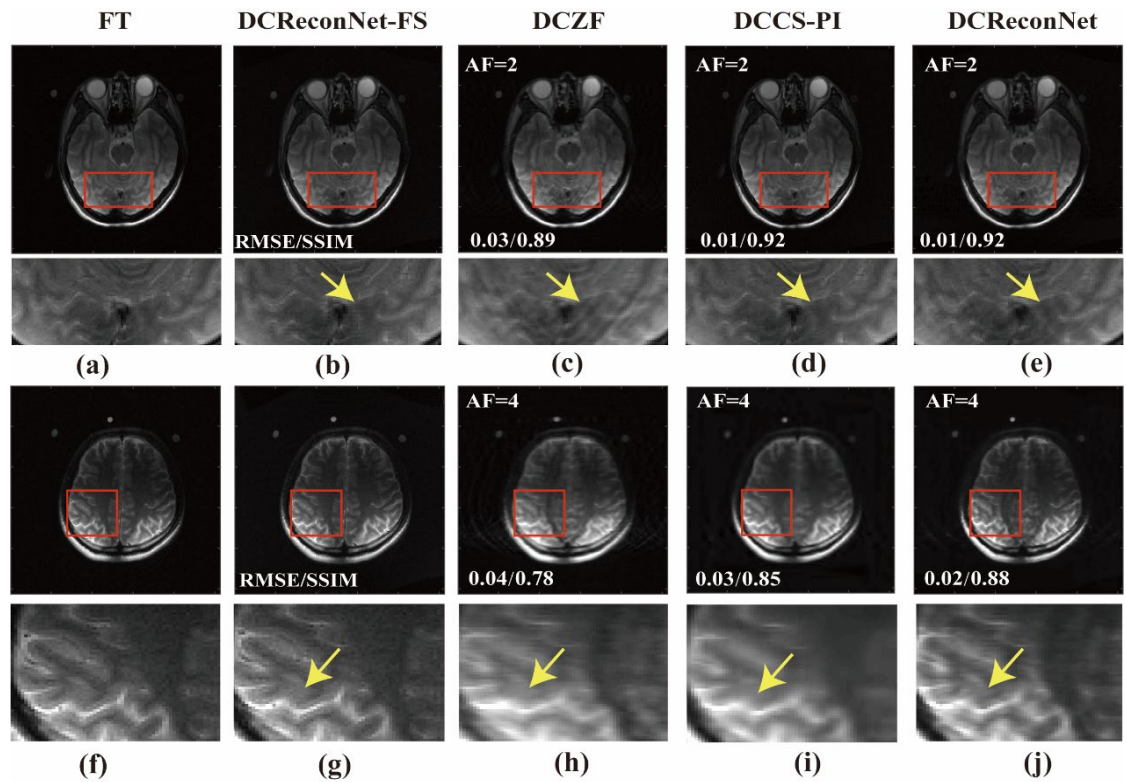

Figure 8. Volunteer brain images acquired with a 6-channel head coil on our experimental MRI-Linac. Reconstructions of fully sampled data with FT (a, f) and DCReconNet (b, g) techniques are shown. Undersampled reconstruction results are shown for DCZF, DCCS-PI and DCReconNet methods at AF=2 (c-e) and AF=4 (h-j). Images (a-e) are from a slice close to the center of the scanner (y=-27.5 mm), and images (f-j) were acquired at y=77.5mm.

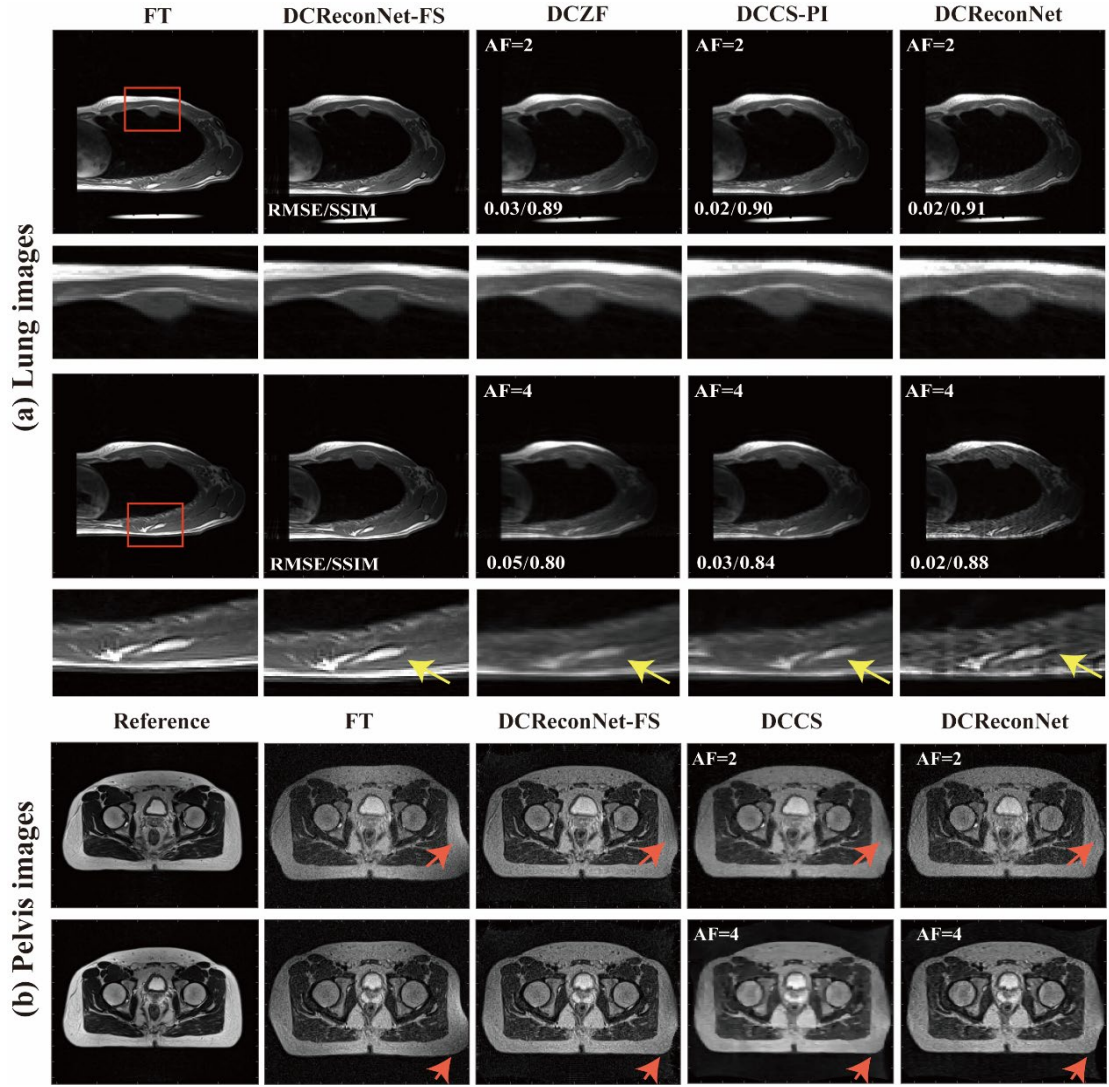

Figure 9. (a) Patient lung images acquired with an 8-channel torso coil on our experimental MRI-Linac. Reconstructions of fully sampled data with FT and DCReconNet techniques are shown in the first two columns. Undersampled reconstruction results are shown for DCZF, DCCS-PI and DCReconNet methods with AF=2 at z = -94mm and AF=4 at z = -82mm. (b) Pelvis images were reconstructed by FT, DCReconNet-FS, DCCS and DCReconNet methods with AF=2 and AF=4, respectively. Fully sampled images acquired by a clinical 3T MRI scanner with vendor's distortion correction were used as references.

4.3 Computational complexity

DCCS, DCCS-PI, UnUnet and DCReconNet methods were implemented on a desktop computer equipped with a Windows 10 Enterprise system and an Intel Xeon CPU @ 3.7GHz and RAM (16GB). Supporting Information Table S1 summarizes the inference processing times with an image resolution of 256 × 256. CS-based reconstructions took 31s and 461s for single-channel and multi-channel acquisitions with CPU, while DCReconNet and UnUNet required only 3s, demonstrating a more than 10-fold reduction in computation time. Additionally, to show the fast reconstruction potential of the neural networks, we executed DCReconNet and UnUNet on a high-performance computer with an Nvidia Tesla V100 P32 GPU, and the latency was 300 ms for both methods, which would facilitate clinical translation.

**Discussion**

Conventional iterative regularization-based algorithms used for MR image reconstruction are computationally slow. In this work, we developed and investigated a deep learning-based fast image reconstruction pipeline, which reconstructs distortion-corrected images from fully sampled and undersampled k-space data. Accelerating the speed of MR image acquisition and reconstruction will be essential to MRI-guided radiotherapy reaching its full clinical potential [49-51]. While techniques such as CS and PI have been deployed with great success to reduce MR acquisition times via undersampling. There is still a need to reduce reconstruction and distortion-correction latencies within 0.5s for online and fast MRI-guided radiation treatments [51-53]. Hence, our results that achieve fast reconstructions with high geometrical precision, compared to traditional regularization methods, are readily applicable in MRI-guided radiotherapy. The residual geometric displacement after using the proposed method was less than 2 mm. Studies [6, 7] have shown that a 2-mm geometric error could lead to ≤5% dosimetry uncertainty, which is considered acceptable for the absorbed dose delivery [54]. The real-time image guidance for radiotherapy treatments requires the total imaging latency including MR acquisition and image reconstruction to be within an acceptable level. Here, the proposed network was tested on retrospectively subsampled data. The clinical in-line implementation of this pipeline, the evaluation on prospectively subsampled data and the MR acquisition latency measurement need to be further investigated.

UnUNet consists of two separate neural networks: a standard unrolling network to reconstruct distorted images and a ResUNet to remove the GNL distortion. The DCReconNet integrates distortion correction into image reconstruction with NUFFT and uses a CNN to learn effective image transformation. As trained on brain data, UnUNet achieves almost comparable performance to the DCReconNet on simulated brain images. However, the proposed DCReconNet has better generalization ability than the UnUNet when tested on other targets (e.g., phantom images) that were not seen in network training. For example, undesired artifacts and image blurring, which were absent in DCReconNet reconstructions, were observed in the UnUNet results. For multi-channel MRI reconstruction, neural networks can be trained with a fixed multi-channel coil setup, and coil sensitivity is included in the training process. However, studies [55] show that networks trained on multi-channel data require re-training for different coil configurations, which will significantly limit their use in practice. In this work, the proposed DCReconNet was trained on single-channel data (i.e., single-channel-input and single-channel-output), which makes it less sensitive to variations of coil configuration. When testing on multi-channel data, the coil-sensitivity-combined images using the sum of squares (SOS) were fed into the network as input so that the coil sensitivity is not included in the network to reduce the computational cost. Nevertheless, the proposed DCReconNet is expected to be applicable to other multi-channel configurations. In the future, possible benefits of leveraging coil-sensitivity profiles need to be further investigated.

Distortion-corrected image reconstruction relies, in part, on accurate GNL field characterization [33, 56-58]. In this work, our previously developed GNLNet [43] was used to provide GNL field information, and studies showed that geometric inaccuracy after distortion correction within the entire ROI was less than the width of one pixel [43, 58]. As GNL fields differ from scanner to scanner, new phantom measurements for GNL field characterization will be required before applying DCReconNet to other systems.

Opportunities exist to apply DCReconNet outside the MRIgRT domain. In particular, we note that there has been significant growth in the number of compact and/or lightweight MRI systems that enable interventional and point-of-care imaging [59, 60]. The compact nature of these systems inherently leads to image distortion due to compromises in gradient design [61]. We believe that such novel systems will increasingly make use of neural network technologies such as DCReconNet to reduce gradient linearity requirement and further shrink the MRI scanner footprint.

In this work, phantom, volunteer brain and patient lung images were acquired with standard Cartesian TSE sequences. Non-Cartesian sequences (e.g., spiral and radial sampling) have recently shown promise for dynamic tumour tracking for radiotherapy [52, 62]. However, eddy currents are a significant additional challenge for non-Cartesian acquisitions, often deviating k-space trajectories and causing significant image artefact [30, 63]. In the future, we will investigate the application of DCReconNet to non-Cartesian reconstruction and distortion-correction problems.

**Conclusion**

In this work, we developed a deep learning-based method to reconstruct distortion-corrected images directly from GNL-corrupted k-space, which was combined with MR accelerating acquisitions including compressed sensing and parallel imaging. Evaluations on the phantom, volunteer brain, pelvis and patient lung images demonstrated that the DCReconNet could better preserve image structural details and significantly improve computational speeds compared with conventional regularization-based reconstruction methods. DCReconNet shows promise for facilitating fast and geometrically precise image guidance for radiotherapy.

**Acknowledgements**

We thank Paul Keall for the valuable discussions. The work was supported by the Australian Government National Health and Medical Research Council (NHMRC) Program (1132471) Grant schemes. P. L. and D. W. receive support from the Cancer Institute of New South Wales (NSW) (fellowships ECF/1032 and ECF/1015). B.W. acknowledges support from the NHMRC Early Career Fellowship scheme (1163010).